\documentclass[aps,showpacs,preprintnumbers,amsmath,amssymb,12pt,floatfix]{revtex4}

\usepackage{latexsym} 
\usepackage{amssymb}  
\usepackage{amsbsy}   
\usepackage{graphicx}       
\usepackage[dvips]{color}
\usepackage{bm}   

\newcommand{\txt}{\textstyle}

\newcommand\eqn[1]{(\ref{#1})}      
\newcommand{\e}{{\rm e}}   

\newcommand{\be}{\begin{equation}}
\newcommand{\ee}{\end{equation}\noindent}
\newcommand{\bea}{\begin{eqnarray}}
\newcommand{\eea}{\end{eqnarray}\noindent}

\newcommand{\half} {{\txt \frac{1}{2}}}

\newcommand{\Det}{\mbox{Det}}

\newcommand{\nn}{\nonumber \\}
\newcommand{\gbar}{\bar\gamma}

\makeatletter 

\def\appendix{\par                              
    \setcounter{section}{0}                     
    \setcounter{subsection}{0}
    \renewcommand{\theequation}{\Alph{section}.\arabic{equation}}
    \renewcommand{\thesection}{Appendix \Alph{section}
                \setcounter{equation}{0}  } 
    \renewcommand{\thesubsection}{\Alph{section}.\arabic{subsection}}
}

\def\applabel#1{\@bsphack
  \protected@write\@auxout{}%
         {\string\newlabel{#1}{{\Alph{section}}{\thepage}}}%
  \@esphack}

\def\section{
\setcounter{equation}{0}        
\@startsection {section}{1}{\z@}{-3.5ex plus -1ex minus 
 -.2ex}{2.3ex plus .2ex}{\large\bf}}
\renewcommand{\theequation}{\arabic{section}.\arabic{equation}}

\def\subsection{\@startsection{subsection}{2}{\z@}{-3.25ex plus -1ex minus 
 -.2ex}{1.5ex plus .2ex}{\normalsize\bf}}

\def\subsubsection{\@startsection{subsubsection}{3}{\z@}{-3.25ex plus
 -1ex minus -.2ex}{1.5ex plus .2ex}{\normalsize}}

\makeatother   

\begin{document}
 
\title{\bf Worldline Instantons II: The Fluctuation Prefactor}

\author{Gerald V.\ Dunne}
\author{Qing-hai Wang} 
\affiliation{Department of Physics, University of Connecticut, 
Storrs, CT 06269-3046, USA}
\author{Holger Gies}
\affiliation{Institut f\"ur Theoretische Physik, Philosophenweg 16, 
69120 Heidelberg, Germany}
\author{Christian Schubert}
\affiliation{Instituto de F\'{\i}sica y Matem\'aticas,
Universidad Michoacana de San Nicol\'as de Hidalgo,
Apdo. Postal 2-82,
C.P. 58040, Morelia, Michoac\'an, M\'exico}


\begin{abstract}
In a previous paper \cite{ds}, it was shown that the worldline expression for the nonperturbative imaginary part of the QED effective action can be approximated by the contribution of a special closed classical path in Euclidean spacetime, known as a {\it worldline instanton}. Here we extend this formalism to compute also the prefactor arising from quantum fluctuations about this classical closed path. We present a direct numerical approach for determining this prefactor, and we find a simple explicit formula for the prefactor in the cases where the inhomogeneous electric field is a function of just one spacetime coordinate. We find excellent agreement between our semiclassical approximation, conventional WKB, and recent numerical results using numerical worldline loops.
\end{abstract}

\begin{titlepage}
\maketitle
\renewcommand{\thepage}{}          

\end{titlepage}

\section{Introduction}
\label{intro}

This paper builds on an earlier paper \cite{ds}, which
presented a new way to compute nonperturbative particle production rates in inhomogeneous electromagnetic fields using a semiclassical approximation to Feynman's worldline path integral formulation of quantum electrodynamics \cite{feynman}. Here we extend this analysis to compute also the subleading prefactor to the particle production rate, by computing the quantum fluctuations about the semiclassical worldline instanton path. The worldline formulation of quantum field theory provides a novel and powerful computational approach to both perturbative and nonperturbative phenomena. Although conceptually as old as relativistic quantum field theory itself, its usefulness for state-of-the-art calculations has been appreciated only in recent years, and largely through its affinity with string methods \cite{gervais,dunbar}. String theory methods have been adapted to quantum field theory, which has inspired many further developments  \cite{halpern,bern,polyakov,tseytlin,strassler,schmidt,mondragon,dhoker,kc,bastianelli} beyond Feynman's original proposal (see \cite{csreview} for a review). In particular, 
the worldline approach provides an efficient way to compute effective actions in
quantum electrodynamics (QED) \cite{rss,cangemi} and quantum chromodynamics (QCD) \cite{sato}.
These effective actions are the generating functionals for scattering amplitudes, and also contain nonperturbative information, for example concerning particle production. This latter aspect is the subject of this current paper. 

The nonperturbative phenomenon of vacuum pair production
\cite{heisenberg,schwinger} has applications in many fields of physics
\cite{greiner,nussinov,kluger,ringwald,roberts,dima,nayak}. It
  is a prominent example of a wider class of quantum-induced nonlinear
  electromagnetic effects \cite{Dittrich:2000zu} which can become accessible
  in current and future strong-field
  experiments \cite{Zavattini:2005tm,heinzl,dipiazza}.
 As is well-known, the QED pair creation by an external field can be
concisely described in terms of the imaginary part of the effective
Lagrangian. Affleck {\it et al} \cite{affleck} studied pair production in a
constant electric field in scalar QED by applying instanton techniques
to the worldline path integral.
This semiclassical worldline approach was generalized in \cite{ds} to the case of {\it inhomogenous} background electric fields. The dominant nonperturbative exponential factor of the imaginary part was computed using special semiclassical worldline loops  called {\it worldline instantons}. Worldline instantons embody the worldline formulation of the conventional field theoretic WKB computations of Brezin and Itzykson \cite{brezin} and Popov {\it et al} \cite{popov}, which were in turn motivated by the pioneering ionization studies of Keldysh \cite{keldysh}. Kim and Page \cite{kimpage} have found an elegant formulation of pair production in the WKB approach using the language of quantum mechanical instantons, and these results are also complementary to our worldline instanton approach. However, these quantum mechanical instantons are not the same as our ``worldline instantons'', which are instantons in the proper-time, rather than in the imaginary time of quantum mechanical tunneling computations. 
Very recently, numerical Monte Carlo techniques have been
developed for the calculation of worldline path integrals \cite{gies,schmidtloop}, and applied to both perturbative and nonperturbative QED processes \cite{giesklingmuller,gisava}. In this paper we find excellent agreement between these numerical results and our semiclassical approximation.

In Section \ref{semiclassical} we discuss the general idea of  worldline instantons as a semiclassical approximation to the nonperturbative part of the worldline effective action. In Section \ref{time} we present a more explicit result for the situation of inhomogeneous electric fields that are just functions of time. Section \ref{space} shows how to extend this to the spatially inhomogeneous case. We conclude with a summary and an outline of possible future work in Section \ref{conclusions}.

\section{Semiclassical Approximation to the Path Integral}
\label{semiclassical}

The Euclidean one-loop effective action for a scalar charged particle (of charge $e$ and mass $m$) in a QED gauge background $A_\mu$ is given
by the worldline path integral expression \cite{csreview}
\bea
\Gamma_{\rm Eucl} [A] =-
\int_0^{\infty}\frac{dT}{T}\, \e^{-m^2T}\!\!\!\!
\int\limits_{x(T)=x(0)} \!\!\!\!\!\! {\mathcal D}x 
\,\, {\rm exp}\left[-\int_0^Td\tau 
\left(\frac{\dot x^2}{4} +i e A\cdot \dot x \right)\right]\quad .
\label{effective}
\eea
Here the functional integral $\int {\mathcal D}x$ is over all closed Euclidean spacetime paths $x^\mu(\tau)$ which are periodic in the proper-time parameter $\tau$, with period $T$. 
We use the path integral normalization conventions of \cite{csreview}. The effective action $\Gamma_{\rm Eucl} [A]$ is a functional of the classical background field $A_\mu(x)$, which is a given function of the space-time coordinates.

If $A_\mu$ corresponds to a Minkowskian electric field, then  the one-loop Minkowski effective action has a non-perturbative imaginary part associated with the pair production from vacuum. This physical interpretation follows from the fact that the Minkowski effective action is related to the vacuum persistence amplitude as
\bea
\langle 0 \, | \, 0 \rangle =e^{i\, \Gamma_{\rm Mink}}\quad.
\label{vacuum}
\eea
An imaginary part of $\Gamma_{\rm Mink}$ is therefore identified with vacuum non-persistence through pair production, such that
\bea
P_{\rm production}=1-e^{-2\, {\rm  Im}\, \Gamma_{\rm Mink}}\approx 2\, {\rm  Im}\, \Gamma_{\rm Mink}\quad.
\label{decay}
\eea
For example, if the background electric field is constant and of magnitude $E$, the leading weak-field expression for this imaginary part is
\bea
{\rm Im}\,\Gamma_{\rm Mink}
\sim V_4^{\rm Mink}\, 
\frac{e^2 E^2}{16\pi^3}
\,\exp\left[-\frac{m^2 \pi}{e E}\right]\quad ,
\label{constant}
\eea
where $V_4^{\rm Mink}$ is the physical (Minkowski) spacetime volume factor.

In \cite{affleck} it was shown how to compute this leading contribution to the imaginary part of $\Gamma_{\rm Mink}$ in the constant field case, by using circular semiclassical paths $x_\mu^{\rm cl}(\tau)$ to approximate the path integral in \eqn{effective} for the Euclidean effective action. In \cite{ds} this idea was generalized to {\it inhomogeneous} electric fields, and the nonperturbative exponential factors were computed in terms of special semiclassical paths called {\it worldline instantons}.  These worldline instantons reduce to the circular paths of Affleck {\it et al} in the limit of a homogeneous background. In this paper we show how to compute also the {\it prefactor} to the exponential factors in the inhomogeneous case, by computing the quantum fluctuations about the worldline instantons. 

In \cite{affleck,ds} the proper-time $T$ integral in \eqn{effective} was done first, followed by the quantum mechanical functional integral, each being evaluated by a steepest descents approximation. However, for inhomogeneous background fields this leads to a {\it nonlocal} fluctuation problem for the functional integral, which makes the prefactor more difficult to compute. So, here we choose instead to evaluate the integrals in \eqn{effective} in the opposite order. That is, we first make a semiclassical approximation to the quantum mechanical path integral in \eqn{effective}, for any $T$, and then evaluate the $T$ integral by steepest descents. It turns out that the form of the worldline instantons remains the same for whichever order of evaluating the integrals, so the results of \cite{ds} form the basis of the computation. The basic idea is that since the worldline formulation is essentially a first-quantized approach, we can use results from the study of semiclassical approximations to quantum mechanical path integrals \cite{levit,kleinertbook}.

Consider the quantum mechanical path integral in \eqn{effective}, with action $S[x]=\int_0^T dt\, L$ for the Lagrangian
\bea
L(x, \dot{x})=\frac{\dot x^2}{4} +i e A\cdot \dot x  \quad .
\label{lag}
\eea
The Euclidean classical Euler-Lagrange equations are
\bea
\ddot{x}_\mu=2ie\, F_{\mu\nu}(x)\,\dot{x}_\nu\quad ,
\label{euler}
\eea
where $F_{\mu\nu}=\partial_\mu A_\nu-\partial_\nu A_\mu$ is the background field strength. Note that it follows immediately that for a classical solution $\dot{x}^2$ is constant:
\bea
\left(\dot{x}^{\rm cl}\right)^2 = a^2 \quad .
\label{a}
\eea
Worldline instantons are periodic solutions to \eqn{euler}, and several classes of explicit solutions were found in \cite{ds}. Furthermore, it was shown in \cite{ds} that these classical worldline instantons straightforwardly determine the nonperturbative exponential factor in ${\rm Im} \,\Gamma_{\rm Mink}$, for inhomogeneous background electric fields.

To compute the prefactor contribution, we need to compute the fluctuations about a worldline instanton. To do this, we need to specify precisely how we sum over all closed loops. There are two standard approaches \cite{polyakov,gerry,csreview}.  The first, which we choose to follow, is to fix a point on each loop and then allow fluctuations about this loop such that the fluctuations vanish at the fixed point. The location of the fixed point is then integrated over. An alternative approach is to consider periodic fluctuations, with the center-of-mass of the loop being kept fixed, and then integrated over at the end. These two approaches are known to give equivalent results after integrating over the loop position \cite{fliegner}. We have further verified that these two approaches also give the same results using worldline instantons, but we found the first approach to be somewhat simpler to implement. Thus we write \eqn{effective} more explicitly as
\bea
\Gamma_{\rm Eucl} [A] =-
\int_0^{\infty}\frac{dT}{T}\, \e^{-m^2T}\int d^4x^{(0)}
\!\!\!\!\!\!\! \int\limits_{x(T)=x(0)=x^{(0)}}\!\!\!\!\!\!\!\!\!\! {\mathcal D}x 
\,\, {\rm exp}\left[-\int_0^Td\tau 
\left(\frac{\dot x^2}{4} +i e A\cdot \dot x \right)\right] .
\label{eff}
\eea
We expand all paths in the functional integral as
\bea
x_{\mu}(\tau)&=&x_{\mu}^{\rm cl}(\tau) + \eta_{\mu}(\tau)\quad , \nn
\eta_\mu(0)&=&\eta_\mu(T)=0\quad .
\label{expansion}
\eea
The first order expansion of the action vanishes by virtue of the Euler-Lagrange equations, and the quadratic term defines the so-called ``secondary action'' \cite{morse,levit}:
\begin{eqnarray}
\delta^2 S [\eta]  &=& \int_0^T d\tau\, \eta_\mu \Lambda_{\mu\nu} \eta_\nu \quad,
\end{eqnarray}
where the fluctuation operator $\Lambda_{\mu\nu}$ has the following general form 
\be
\Lambda_{\mu\nu}\equiv -\frac{1}{2}\,\delta_{\mu\nu}\, \frac{d^2}{d\tau^2} -\frac{d}{d\tau}\,Q_{\nu\mu}+Q_{\mu\nu}\, \frac{d}{d\tau}+R_{\mu\nu}\quad,
\ee
with
\bea
Q_{\mu\nu}&\equiv& \frac{\partial^2 L}{\partial x_\mu \partial \dot{x}_\nu}\quad,\nn
R_{\mu\nu}&\equiv& \frac{\partial^2 L}{\partial x_\mu \partial x_\nu} \quad .
\label{flucop}
\eea
The equations of motion for the fluctuations are known as the Jacobi equations \cite{morse}:
\bea
\Lambda_{\mu\nu}\, \eta_\nu=0\quad .
\label{jacobi}
\eea
The semiclassical approximation to quantum mechanical path integrals  \cite{levit,kleinertbook} leads to the following simple result:
\bea
\int\limits_{x(T)=x(0)=x^{(0)}} \!\!\!\!\!\!\!\!{\mathcal D}x 
\,\, {\rm exp}\left[-\int_0^Td\tau 
\left(\frac{\dot x^2}{4} +i e A\cdot \dot x \right)\right]
\approx \frac{e^{i\,\theta}e^{-S[x^{\rm cl}](T)}}{(4\pi T)^2} \sqrt{\frac{\left | \det\left[\eta_{\mu,\, {\rm free}}^{(\nu)}(T)\right] \right |}{\left | \det\left[\eta_\mu^{(\nu)}(T)\right] \right |}} .
\label{semi}
\eea
The important result from \cite{levit,kleinertbook} is that the determinant $\det\left[\eta_\mu^{(\nu)}(T)\right]$ is simply a {\it finite dimensional} $4\times 4$ determinant, formed from solutions, $\eta_\mu^{(\nu)}(\tau)$, to the Jacobi equations \eqn{jacobi} with the {\it initial value boundary conditions}
\bea
\eta_\mu^{(\nu)}(0)=0\quad ; \quad \dot{\eta}_\mu^{(\nu)}(0)=\delta_{\mu\nu} \quad , \quad (\mu, \nu = 1, 2, 3, 4),
\label{iv}
\eea
and evaluated at the endpoint $\tau=T$. Given the worldline instanton, $x^{\rm cl}(\tau)$, it is straightforward to implement this determinant computation numerically. Similarly, the free case is just given by the corresponding equations with $\Lambda_{\mu\nu}^{\rm free}\equiv -\frac{1}{2}\,\delta_{\mu\nu}\, \frac{d^2}{d\tau^2}$, in which case $\det\left[\eta_{\mu,\, {\rm free}}^{(\nu)}(T)\right] =T^4$, with these normalizations. In the next Section we show that for certain classes of inhomogeneous background fields it is possible to be even more explicit and find a simple analytic expression for this determinant in terms of the classical worldline instanton paths $x^{\rm cl}$ themselves. The (constant) phase factor $e^{i\theta}$ in \eqn{semi} is determined by the Morse index \cite{morse,levit} of the operator $\Lambda$, which counts the number of times  $\det\left[\eta_\mu^{(\nu)}(\tau)\right]$ vanishes in the interval from $0$ to $T$. In the cases considered in this paper we find this phase factor reduces to $\pm 1$, as described below.

\section{Time-dependent electric field}
\label{time}

In this section we illustrate our semiclassical procedure in a class
of models where the computation can be done very explicitly. These
cases \cite{ds} are those where the electric field points in a given
direction in space (say the $x_3$ direction), and is {\it either} (i)
a function only of time, {\it or}  (ii) a function only of $x_3$. The
case (i) has been widely studied in conventional WKB
\cite{brezin,popov}, while the case (ii) has been studied using WKB
instantons \cite{kimpage}, and numerically using worldline loops
\cite{giesklingmuller}. We first consider case (i), that of a
time-dependent electric field, where the imaginary Euclidean
gauge field (corresponding to a real Minkowski electric field)
can be written as 
\begin{equation}
A_3(x_4)=-i\frac{E}{\omega}f(\omega\, x_4)\quad .
\label{tgauge}
\end{equation}
Throughout the paper we will illustrate the method explicitly using the examples of $f(\omega x_4)=\tan(\omega x_4)$, corresponding to a single-pulse Minkowski electric field $E(t)=E\, {\rm sech}^2(\omega t)$, and $f(\omega x_4)=\sinh(\omega x_4)$, corresponding to a periodic Minkowski electric field $E(t)=E\, \cos(\omega t)$.
Motivated by the analogy with Keldysh's classic work \cite{keldysh} on atomic ionization, the inhomogeneity of the background is usually characterized by the dimensionless {\it adiabaticity parameter}
\bea
\gamma\equiv \frac{m\omega}{e E}\quad .
\label{g}
\eea
\subsection{Classical solutions}
The classical  equations of motion \eqn{euler} become
\begin{eqnarray}
\ddot{x}_1 &=& 0 \quad,\nn
\ddot{x}_2 &=& 0 \quad,\nn
\ddot{x}_3 &=& -2 eE f^\prime(\omega x_4)\, \dot{x}_4 \quad,\nn
\ddot{x}_4 &=& 2 eE f^\prime(\omega x_4)\, \dot{x}_3 \quad.
\label{teuler}
\end{eqnarray} 
For periodic solutions, $x_1^{\rm cl}$ and $x_2^{\rm cl}$ must be constant, so the relation \eqn{a} reduces to
\begin{equation}
(\dot{x}_3^{\rm cl})^2+(\dot{x}_4^{\rm cl})^2\equiv a^2\quad ,
\label{a2}
\end{equation}
and we are left with an effectively two-dimensional problem in the $(x_3, x_4)$ plane. 
The third equation in \eqn{teuler} can be integrated immediately:
\begin{equation}
\dot{x}_3^{\rm cl}=-\frac{2eE}{\omega}f(\omega x_4^{\rm cl})\quad .
\label{x3dot}
\end{equation}
Here we have chosen the integration constant to vanish in order to have a periodic solution.
Using \eqn{a2} we find that the remaining equation is a first-order nonlinear equation for $x_4^{\rm cl}$:
\begin{equation}
\dot{x}_4^{\rm cl}=\pm a \sqrt{1-\left(\frac{ f(\omega x_4^{\rm cl})}{\gbar}\right)^2}
\quad ,
\label{x4dot}
\end{equation}
where $\gbar$ is defined as
\bea
\gbar\equiv \frac{a\omega}{2 eE}=\frac{a}{2m}\,\gamma\quad .
\label{gammabar}
\eea
The nonlinear equation \eqn{x4dot} is the same as the one considered in \cite{ds}, with $\gbar$ in place of $\gamma$, so we can use the results of \cite{ds} to write the explicit form of the solutions $(x_3^{\rm cl}, x_4^{\rm cl})$. For example:

\begin{itemize}

\item For the Minkowski electric field $E(t)=E\, {\rm sech}^2(\omega t)$, we have  $f(\omega x_4)=\tan (\omega x_4)$, and the worldline instanton loop is \cite{ds}:
\bea
x_3^{\rm cl}(\tau)&=&\frac{1}{\omega}\,\frac{1}{\sqrt{1+\gbar^2}} \, {\rm arcsinh}\left[\gbar\, \cos\left(2 e E \sqrt{1+\gbar^2}\,(\tau+\tau_0)\right)\right] \quad , \nn 
x_4^{\rm cl}(\tau)&=&\frac{1}{\omega}\,  \arcsin\left[\frac{\gbar}{\sqrt{1+\gbar^2}}\, \sin\left(2 e E \sqrt{1+\gbar^2}\, (\tau+\tau_0)\right)\right] \quad .
\label{psols-sech}
\eea
Periodicity of this solution enforces the following functional relation between $T$ and $\gbar$ (or, equivalently, between $T$ and $a$):
\bea
T=\frac{\pi}{e E}\, \frac{1}{\sqrt{1+\gbar^2}} \quad .
\label{period-sech}
\eea

\item For the Minkowski electric field $E(t)=E\, \cos(\omega t)$, we have  $f(\omega x_4)=\sinh(\omega x_4)$, and the worldline instanton loop is \cite{ds}:
\bea
x_3^{\rm cl}(\tau)&=&\frac{1}{\omega}\, {\rm arcsin}\left[\frac{\gbar}{\sqrt{1+\gbar^2}}\, {\rm cd}\left(2 e E \sqrt{1+\gbar^2}\,  (\tau+\tau_0)\, {\Bigg |}  \frac{\gbar^2}{1+\gbar^2}\right)\right]\quad ,
\nn 
x_4^{\rm cl}(\tau)&=&\frac{1}{\omega}\, {\rm arcsinh}\left[\frac{\gbar}{\sqrt{1+\gbar^2}}\, {\rm sd}\left(2 e E \sqrt{1+\gbar^2}\,  (\tau+\tau_0)\, {\Bigg |}  \frac{\gbar^2}{1+\gbar^2}\right)\right] \quad .
\label{psols-cos}
\eea
Here, cd and sd are Jacobi elliptic functions \cite{abramowitz}. Periodicity of this solution enforces the following functional relation between $T$ and $\gbar$ (or, equivalently, between $T$ and $a$):
\bea
T=\frac{2}{e E}\, \frac{{\bf K}\left(\frac{\gbar^2}{1+\gbar^2}\right)}{\sqrt{1+\gbar^2}} \quad ,
\label{period-cos}
\eea
where  ${\bf K}(\alpha)$ denotes the elliptic quarter period \cite{abramowitz}. 

\end{itemize}
In these classical solutions, $\tau_0$ is an arbitrary constant whose physical interpretation is to label the fixed point on the loop about which the fluctuations are taken, as described above. The integration over this collective coordinate is discussed below [see \eqn{collective}].

An interesting geometric observation is that the curvature of the planar loop $(x_3^{\rm cl}(\tau), x_4^{\rm cl}(\tau))$ at any given point on the curve is
\be
\kappa(\tau)= \frac{\dot{x}_3\, \ddot{x}_4-\dot{x}_4\, \ddot{x}_3}{\left(\dot{x}_3^2+\dot{x}_4^2\right)^{3/2}} 
= \frac{2e}{a}\, E\, f^\prime (\omega x_4) \quad ,
\label{curvature}
\ee
which is proportional to the Euclidean electric field strength $E f^\prime(\omega x_4)$ evaluated at that point. Thus, for the constant electric field case the classical path is a circle, which has constant curvature, while for an inhomogeneous electric field the planar curvature changes along the loop, in such a way that it tracks the electric field. See \cite{formiga} for a recent investigation of the connection between electromagnetic fields and the geometry of the associated particle trajectories.

\subsection{Fluctuation operator and its determinant}

For the time-dependent fields of the form \eqn{tgauge}, the fluctuation operator \eqn{flucop} can be restricted to its components in the $(x_3, x_4)$ plane :
\bea
\Lambda &=& \left(\begin{array}{cc}
- \half \frac{d^2}{d^2\tau} & - eE f^\prime(\omega x_4^{\rm cl})\frac{d}{d\tau} -  eE\omega f^{\prime\prime}(\omega x_4^{\rm cl})\, \dot{x}_4^{\rm cl}\\
eE f^\prime(\omega x_4^{\rm cl})\frac{d}{d\tau} & - \half \frac{d^2}{d^2\tau} + eE \omega f^{\prime\prime}(\omega x_4^{\rm cl})\,\dot{x}_3^{\rm cl}
\end{array}
\right)\nn
&=& \frac{1}{2}\left(\begin{array}{cc}
- \frac{d^2}{d^2\tau}& \frac{\ddot{x}_3^{\rm cl}}{\dot{x}_4^{\rm cl}}\frac{d}{d\tau} +  \frac{d}{d\tau}\left(\frac{\ddot{x}_3^{\rm cl}}{\dot{x}_4^{\rm cl}}\right)\\
- \frac{\ddot{x}_3^{\rm cl}}{\dot{x}_4^{\rm cl}}\frac{d}{d\tau} & - \frac{d^2}{d^2\tau} - \frac{\dot{x}_3^{\rm cl}}{\dot{x}_4^{\rm cl}} \frac{d}{d\tau}\left(\frac{\ddot{x}_3^{\rm cl}}{\dot{x}_4^{\rm cl}}\right)
\end{array}
\right) \quad .
\label{tfluc}
\eea
To compute the fluctuation determinant we use the semiclassical quantum mechanical path integral result \eqn{semi}. Thus, we need to find solutions to the Jacobi equations $\Lambda\,\eta=0$ in \eqn{jacobi}, satisfying the initial value conditions \eqn{iv}. Remarkably, for the fluctuation operator \eqn{tfluc} we can write all four independent solutions to the Jacobi equations \eqn{jacobi}:
\begin{eqnarray}
\phi^{(1)}(\tau) &=& \left(\begin{array}{c}
1\\ 
0\end{array}\right) \quad,\nn
\phi^{(2)}(\tau) &=& \left(\begin{array}{c}
\dot{x}_3^{\rm cl}(\tau)\\ 
\dot{x}_4^{\rm cl}(\tau)\end{array}\right) \quad,\nn
\phi^{(3)}(\tau) &=& \left(\begin{array}{c}
\dot{x}_3^{\rm cl}(\tau)  \int_0^{\tau}dt\, \frac{1}{\left[\dot{x}_4^{\rm cl}(t)\right]^2} -  \int_0^{\tau}dt\, \frac{\dot{x}_3^{\rm cl}(t)}{\left[\dot{x}_4^{\rm cl}(t)\right]^2}\\ 
\dot{x}_4^{\rm cl}(\tau)  \int_0^{\tau}dt\, \frac{1}{\left[\dot{x}_4^{\rm cl}(t)\right]^2}\end{array}\right) \quad,\nn
\phi^{(4)}(\tau) &=& \left(\begin{array}{c}
\dot{x}_3^{\rm cl}(\tau)  \int_0^{\tau}dt\, \frac{\dot{x}_3^{\rm cl}(t)}{\left[\dot{x}_4^{\rm cl}(t)\right]^2} -  a^2\int_0^{\tau}dt\, \frac{1}{\left[\dot{x}_4^{\rm cl}(t)\right]^2}\\ 
\dot{x}_4^{\rm cl}(\tau)  \int_0^{\tau}dt\, \frac{\dot{x}_3^{\rm cl}(t)}{\left[\dot{x}_4^{\rm cl}(t)\right]^2}\end{array}\right) \quad .
\label{zeromodes}
\end{eqnarray}
The first zero mode $\phi^{(1)}$ corresponds to translational invariance in the $x_3$ direction, while the second zero mode $\phi^{(2)}$ corresponds to invariance under shifts of the starting point on the loop. The third and the fourth zero modes $\phi^{(3,4)}$ are associated with the velocity whose magnitude $a$ is a constant.  The linear combinations satisfying the initial conditions \eqn{iv} are
\begin{eqnarray}
\eta^{(3)}(\tau) &=& \dot{x}_3^{\rm cl}(0)\phi^{(3)}(\tau)-\phi^{(4)}(\tau)\quad, \nn
\eta^{(4)}(\tau) &=& \dot{x}_4^{\rm cl}(0)\phi^{(3)}(\tau) \quad .
\end{eqnarray}
A simple computation shows that the fluctuation determinant is 
\bea
\Det (\Lambda) &\equiv & \det \left[\eta^{(3)}(T), \,\eta^{(4)} (T)\right]\nn
& =& \frac{\left[\dot{x}_4^{\rm cl}(0)\right]^3}{ \dot{x}_4^{\rm cl}(T)} \left[a^2 I_1^2(T)-I_2^2(T)\right] \quad,
\label{det1}
\eea
where
\begin{eqnarray}
I_1(\tau) &\equiv& \frac{\dot{x}_4^{\rm cl}(\tau)}{\dot{x}_4^{\rm cl}(0)}\, \int_0^\tau d t\,  \frac{1}{\left[\dot{x}_4^{\rm cl}(t)\right]^2} \quad,\nn
I_2(\tau) &\equiv&  \frac{\dot{x}_4^{\rm cl}(\tau)}{\dot{x}_4^{\rm cl}(0)}\, \int_0^\tau d t\,  \frac{\dot{x}_3^{\rm cl}(t)}{\left[\dot{x}_4^{\rm cl}(t)\right]^2} \quad .
\end{eqnarray}
For example, for the two special cases considered above, we find
\begin{itemize}
\item For the Minkowski electric field $E(t)=E\, {\rm sech}^2(\omega t)$, with  $f(\omega x_4)=\tan (\omega x_4)$, the classical velocities (with the periodicity condition \eqn{period-sech} imposed) are \cite{ds}:
\bea
\dot{x}_3^{\rm cl}(\tau)&=&- a\, \frac{\sin\left(\frac{2\pi}{T}\, (\tau+\tau_0)\right)}{\sqrt{1+\gbar^2 \cos^2\left(\frac{2\pi}{T} \, (\tau+\tau_0)\right)}}
\quad,\nn 
\dot{x}_4^{\rm cl}(\tau)&=&a\,\sqrt{1+\gbar^2}\, \frac{\cos (\frac{2\pi}{T}\, (\tau+\tau_0))}
{\sqrt{1+\gbar^2 \cos^2\left(\frac{2\pi}{T}\, (\tau+\tau_0)\right)}} \quad .
\label{vel-sech}
\eea
Then the integrals $I_1(T)$ and $I_2(T)$ in this case can be evaluated as
\bea
I_1(T)&=& \left(\frac{\omega}{2eE}\right)^2 \frac{T}{1+\gbar^2(T)}\quad,\nn
I_2(T)&=&0 \quad .
\label{sech-is}
\eea
\item For the Minkowski electric field $E(t)=E\, \cos(\omega t)$, with  $f(\omega x_4)=\sinh(\omega x_4)$, the classical velocities  (with the periodicity condition \eqn{period-cos} imposed) are \cite{ds}:
\bea
\dot{x}_3^{\rm cl}(\tau)&=&-a\, \frac{1}{\sqrt{1+\gbar^2}}\, {\rm sd}\left[\frac{4}{T}\, {\bf K}\left(\frac{\gbar^2}{1+\gbar^2}\right) (\tau+\tau_0)\, {\Bigg |}  \frac{\gbar^2}{1+\gbar^2}\right]
\quad,\nn 
\dot{x}_4^{\rm cl}(\tau)&=&a\, {\rm cd}\left[\frac{4}{T}\, {\bf K}\left(\frac{\gbar^2}{1+\gbar^2}\right) (\tau+\tau_0)\, {\Bigg |}  \frac{\gbar^2}{1+\gbar^2}\right] \quad .
\label{vel-cos}
\eea
Then the integrals $I_1(T)$ and $I_2(T)$ in this case can be evaluated as
\bea
I_1(T)&=&\left(\frac{\omega}{2 eE}\right)^2 \, T\, \left[\frac{{\bf K}\left(\frac{\gbar^2(T)}{1+\gbar^2(T)}\right)-{\bf E}\left(\frac{\gbar^2(T)}{1+\gbar^2(T)}\right)}{{\gbar^2(T)\bf K}\left(\frac{\gbar^2(T)}{1+\gbar^2(T)}\right)}\right] \quad,\nn
I_2(T)&=&0 \quad .
\label{cos-is}
\eea
\end{itemize}
The determinant \eqn{det1} can be simplified using periodicity, which implies $\dot{x}_4^{\rm cl}(0)=\dot{x}_4^{\rm cl}(T)$, and the vanishing of $I_2(T)$. Therefore, we find the simple expression for the fluctuation determinant:
\bea
\Det (\Lambda) = \left[\frac{2eE}{\omega}\,\dot{x}_4^{\rm cl}(0)\, \gbar(T)\, I_1(T)\right]^2 \quad,
\label{det2}
\eea
where we have written $a=\frac{eE}{\omega}\gbar$ in terms of $\gbar(T)$ to stress that the periodicity condition fixes the parameter $\gbar$ to be a particular function of $T$, as in \eqn{period-sech} and \eqn{period-cos}, for example.

Now recall that we still need to evaluate the $4$-dimensional space-time integral over the fixed point on the closed loops:
\begin{eqnarray}
\int d^4 x^{(0)}\equiv\int dx_1(0)\, dx_2(0)\, dx_3(0)\, dx_4(0) = V_3 \int d\tau_0\, \dot{x}_4^{\rm cl}(0) \quad ,
\label{collective}
\end{eqnarray}
where $V_3$ is the $3$-space volume. Observe that this factor of $\dot{x}_4^{\rm cl}(0)$ cancels against the same factor in $\sqrt{\Det(\Lambda)}$, so that the spacetime integration effectively contributes a 3-volume factor $V_3$, and a factor of $\frac{T}{2}$. This last factor is just the collective coordinate contribution arising from invariance under shifts of the starting point on the loop, which gives rise to the second of the zero modes in \eqn{zeromodes}.

Finally, to fix the phase factor in \eqn{semi} we need the Morse index of the fluctuation operator $\Lambda$. This can be evaluate either as the number of negative eigenvalues of the operator $\Lambda$, or as the number of times $\det\left[\eta_\mu^{(\nu)}(\tau)\right]$ vanishes in the interval from $0$ to $T$. We find that for the time dependent fields of the form discussed here, the Morse index is $2$, leading to a phase factor $e^{-i 2\pi/2}=-1$. Thus, collecting all the pieces, we see that the semiclassical approximation \eqn{semi} to the quantum mechanical path integral leads to :
\begin{eqnarray}
\Gamma_{\rm Eucl}^{\rm semi}  \approx V_3 \, \frac{1}{(4\pi)^2}\,\frac{\omega}{4eE}\,\int_0^{\infty}\frac{dT}{T} \, \frac{\exp\left[-\left(S[x^{\rm cl}](T)+m^2 T\right)\right]}{\gbar(T)I_1(T)} \quad .
\label{tint}
\end{eqnarray}

\subsection{Extracting the Nonperturbative Imaginary Part}

So far we have been computing the Euclidean effective action, using the worldline expression \eqn{effective}. We relate this to the imaginary part of the  physical Minkowski effective action according to the following conventions:
\bea
e^{i\, S_{\rm Mink}}=e^{i\int dt\, L_{\rm Mink}}=e^{\int dy_4\, L_{\rm Mink}}=e^{-S_{\rm Eucl}} =e^{-\int dy_4 L_{\rm Eucl}}\quad .
\eea
Thus we identify $y_4=i\, t$, $L_{\rm Mink}=- L_{\rm Eucl}$, and 
\bea
\Gamma_{\rm Mink}=i\, \Gamma_{\rm Eucl}\quad .
\label{mink}
\eea
Some care is needed for extracting the desired Minkowski imaginary
part:
\bea
{\rm Im}\, \Gamma_{\rm Mink}&=&  {\rm Im} \int d^3y\, \int dt\,\,
  {\mathcal L}_{\rm Mink}(t,\vec y)\nn
&=& - {\rm Im} \int d^3y\, \int dt\,\,
  {\mathcal L}_{\rm Eucl}(i t,\vec y)\nn 
&=&  {\rm Re}  \int d^3y\,\,\, i \int dt\,\,
  {\mathcal L}_{\rm Eucl}(i t,\vec y)\nn 
&=&  {\rm Re}  \int d^3y\,  \int_{\mathcal C} dy_4\,\,
  {\mathcal L}_{\rm Eucl}(y_4,\vec y)\nn 
&=&{\rm Re}\, \Gamma_{\rm Eucl,\mathcal C} \quad ,
\label{lcf}
\eea
where the contour $\mathcal C$ goes along the imaginary axis from
$-i\infty$ to $i\infty$. For instance, for time-independent electric
background fields, we obtain
\begin{equation}
\text{Im}\Gamma_{\text{Mink}}\equiv \int dt\, \text{Im}\,
L_{\text{Mink}} =\text{Re}\, \int_{\mathcal C} dy_4 \, L_{\text{Eucl}} 
=-\int dt\, \text{Im}\, L_{\text{Eucl}}, \label{eq:imexconst}
\end{equation}
such that the imaginary part of the Minkowski Lagrange function is
given by (minus) the imaginary part of the Euclidean Lagrange
function (note that $L=\int d^3y\, \mathcal L$). 

On the other hand, for time-dependent electric fields for which the
$y_4$ contour $\mathcal C$ can be rotated onto the real axis, we find
\begin{equation}
\text{Im}\Gamma_{\text{Mink}}=\text{Re}\, \int_{\mathcal C} dy_4 \,
L_{\text{Eucl}}(y_4) =\text{Re}\, \int_{-\infty}^\infty dy_4\,
L_{\text{Eucl}}(y_4) =\text{Re}\, \Gamma_{\text{Eucl}}.
\label{eq:imexrotcontour}
\end{equation}
For the electric fields considered in the present work, the contour
can indeed be rotated, since worldline instantons extending to $y_4\to
\pm\infty$ yield a vanishing contribution that drops off sufficiently
fast at complex infinity. 
Thus to find the nonperturbative imaginary part of the Minkowski
effective action in the semiclassical approximation, we need to
extract the {\it real part} of the propertime integral expression in
\eqn{tint}.

\subsection{The $T$ integral}

In general, the $T$ integral in the semiclassical expression \eqn{tint} cannot be done analytically. However, in the weak field limit the physical nonperturbative part may be extracted directly using a steepest descents approximation, by evaluating the $T$ integral in the vicinity of a critical point. To do so, we study the 
exponent in \eqn{tint} :
\begin{equation}
\Delta(T) =  S[x^{\rm cl}](T)+m^2 T\quad .
\label{exponent}
\end{equation}
This notation emphasizes the fact that the action $S[x^{\rm cl}]$, evaluated on the worldline instanton path $x^{\rm cl}(\tau)$, is a function of $T$. Before considering the general case, we illustrate with the example of the Minkowski electric field $E(t)=E\, {\rm sech}^2(\omega\, t)$. Then, using the worldline instanton $x_\mu^{\rm cl}(\tau)$ found in \eqn{psols-sech}, we find
\bea
\Delta(T)&=&\frac{\pi^2}{\omega^2T}\left(1-\frac{eET}{\pi}\right)^2+m^2T \nn
&=& \frac{m^2\pi}{e E}\left[ \frac{e E T}{\pi}\left(\frac{1+\gamma^2}{\gamma^2}\right) +\frac{1}{\gamma^2}\frac{\pi}{e E T}-\frac{2}{\gamma^2}\right]\quad,
\label{delta-sech}
\eea
where $\gamma$ is defined in \eqn{g}. And we have used the relation \eqn{period-sech} between $T$ and $\gbar$, which follows from the periodicity of the solution. This can be written as
\bea
\gbar(T)=\frac{\pi}{e E T}\, \sqrt{1-\left(\frac{e E T}{\pi}\right)^2 } \quad .
\label{gbar-sech}
\eea
Using this, we can express $I_1(T)$ in \eqn{sech-is} directly as a function of $T$:
\bea
I_1(T)=\left(\frac{\omega}{2\pi}\right)^2 T^3\quad .
\label{sech-newi1}
\eea
Thus,
\bea
\Gamma_{\rm Eucl}^{\rm semi} \approx V_3 \, \frac{1}{16 \pi \omega}\,\int_0^{\infty}\frac{dT}{T^3\, \sqrt{1-\left(\frac{e E T}{\pi}\right)^2 }} \, \exp\left\{-\frac{m^2\pi}{e E}\left[ \frac{e E T}{\pi}\left(\frac{1+\gamma^2}{\gamma^2}\right) +\frac{1}{\gamma^2}\frac{\pi}{e E T}-\frac{2}{\gamma^2}\right]\right\}\quad .\nn
\label{tint-sech}
\eea
Observe that there is a branch cut along part of the real $T$ axis, so the integral has both a real and an imaginary part. As explained in the previous section, to obtain the physical nonperturbative imaginary part of the Minkowski effective action we need the real part of \eqn{tint-sech}. The branch point occurs at 
\bea
T_b=\frac{\pi}{e E} \quad .
\label{branch}
\eea
The imaginary part of \eqn{tint-sech} comes from an integral across the cut, which extends from $T_b$ to infinity, but we are instead interested in the contribution from the region to the left of the branch point. Observe that it is natural to rescale $T$ as $\frac{e E T}{\pi}$, so that in the weak field limit $E\to 0$, we expect a dominant contribution from the vicinity of critical points of $\Delta(T)$. From \eqn{delta-sech} we find the critical point of the exponent is at
\bea
T_c=\frac{\pi}{e E}\,\frac{1}{\sqrt{1+\gamma^2}} \quad ,
\label{critical-sech}
\eea
which falls to the left of the branch point, for all frequencies $\omega$ ({\it i.e.}, all $\gamma$). So this critical point produces the required dominant contribution. Comparing \eqn{period-sech} and \eqn{critical-sech}, we see that the critical point occurs when 
\bea
\gbar(T_c)=\gamma \quad .
\label{special}
\eea
Recalling the definition \eqn{gammabar} of $\gbar$, we see that the critical point occurs when $a=2m$, which is precisely the value used in \cite{ds} when the propertime  $T$ integral was done first. In fact, this property is general, as we show in the next section. 

At the critical point, the exponent is
\bea
\Delta(T_c)=\frac{m^2\pi}{e E} \, \left(\frac{2}{1+\sqrt{1+\gamma^2}} \right) \quad .
\eea
The critical point is a minimum since
\bea
\Delta^{\prime\prime}(T_c)=\frac{2 e Em^2}{\pi}\, \frac{\left(1+\gamma^2\right)^{3/2}}{\gamma^2}\quad ,
\eea
which is positive. Thus, we can use Laplace's method to approximate the integral in the weak field limit, leading to:
\bea
\left[ {\rm Im}\,\Gamma_{\rm Mink}^{\rm semi}\right]_{E(t)=E\, {\rm sech}^2(\omega t)}  &\approx &
 V_3^{\rm Mink} \, \frac{1}{(4\pi)^2}\,\frac{\omega}{4eE}\, \sqrt{\frac{2\pi}{\Delta^{\prime\prime}(T_c)} }  \, \frac{e^{-\Delta(T_c)}}{T_c \gbar(T_c)\, I_1(T_c)}
 \label{laplace-sech} \\
&=& V_3^{\rm Mink}\, \frac{\left(e E\right)^{5/2}}{16\pi^3m\omega} \left(1+\gamma^2\right)^{5/4}\, \exp\left[-\frac{m^2\pi}{e E} \, \left(\frac{2}{1+\sqrt{1+\gamma^2}} \right) \right]\,\, .
\nonumber
\eea
This should be compared to the {\it locally constant field (LCF) approximation} in \eqn{lcf}:
\bea
\left[ {\rm Im}\,\Gamma_{\rm Mink}^{\rm LCF}\right]_{E(t)=E\, {\rm sech}^2(\omega t)}  &\approx &
V_3^{\rm Mink}\, \frac{e^2 E^2}{16\pi^3}\, \int_{-\infty}^\infty dt\, {\rm sech}^4(\omega t)\, \exp\left[-\frac{m^2\pi}{e E}\,\cosh^2(\omega t)\right]\nn
&\sim& V_3^{\rm Mink}\, \frac{\left(e E\right)^{5/2}}{16\pi^3m\omega} \, \exp\left[-\frac{m^2\pi}{e E}  \right]\quad .
\label{lcf-sech}
\eea
Here, in the second line of \eqn{lcf-sech} we have made the same weak field limit approximation as was made in \eqn{laplace-sech}.
Then \eqn{lcf-sech} agrees perfectly with the static limit ($\gamma\to 0$) of our semiclassical result \eqn{laplace-sech}.
The semiclassical result \eqn{laplace-sech} also agrees with Popov's WKB result \cite{popov} for this time dependent electric field, and with the Borel analysis of the resummed derivative expansion \cite{dunnehall}. Note that the temporal inhomogeneity of the field enhances the local pair production rate, as discussed in \cite{popov,ds}.

\subsection{General case}

We now show that the example given in the previous section is quite general for time dependent electric fields of the form \eqn{tgauge}. Using the classical equations of motion \eqn{teuler}, we can write
\bea
S[x^{\rm cl}](T)&=& \int_0^T d\tau\,\left\{\frac{1}{4}\left[\left(\dot{x}_3^{\rm cl}\right)^2+\left(\dot{x}_4^{\rm cl}\right)^2\right]+\frac{e E}{\omega}\, f\left(\omega x_4^{\rm cl}\right)\, \dot{x}_3^{\rm cl}\right\}\nn
&=&-\frac{a^2}{4}\, T +\frac{1}{2}\,\int^T_0\, d\tau\, \left(\dot{x}_4^{\rm cl}\right)^2 \quad .
\eea
Thus, the exponent $\Delta(T)$ can be expressed as a (complicated) function of $T$:
\bea
\Delta(T)&=&m^2 T\left(1-\frac{\gbar^2(T)}{\gamma^2}\right)+\frac{\pi m^2}{e E}\, \frac{\gbar^2(T)}{\gamma^2} \, g(\gbar^2)\quad ,
\label{exponent2}
\eea
where we have defined the important function $g$ :
\bea
g(\gbar^2)&\equiv& \frac{2}{\pi} \int_{-1}^1 \frac{dy\, \sqrt{1-y^2}}{|f^\prime |}\quad .
\label{gfunction}
\eea
In this definition of the function $g$, we have written $y=\frac{f(\omega \, x_4^{\rm cl})}{\gbar}$, and $f^\prime$ means the derivative is re-expressed as a function of $y$. With the same notation, the periodicity condition can be expressed in general as
\bea
T=\frac{1}{e E}\int_{-1}^1 \frac{dy}{|f^\prime |\, \sqrt{1-y^2}}
\equiv \frac{\pi}{e E}\, P(\gbar^2) \quad ,
\label{periodic}
\eea
which determines $\gbar$ as a function of $T$. Note that the two functions, $g$ and $P$ are related as follows:
\bea
P(z)=\frac{d}{d z}\left(z\, g(z)\right)\quad .
\label{pg}
\eea

For example:
\begin{itemize}
\item For the Minkowski electric field $E(t)=E\, {\rm sech}^2(\omega t)$, we have $f(\omega x_4)=\tan(\omega x_4)$, so that 
\be
f^\prime(\omega x_4)= \sec^2(\omega x_4)= 1+\gbar^2\, y^2\quad .
\ee
Then the exponent \eqn{exponent2} involves the function 
\bea
g(\gbar^2)=\frac{2}{\pi}\int_{-1}^1 \frac{dy \, \sqrt{1-y^2}}{(1+\gbar^2 \, y^2)}=\frac{2}{1+\sqrt{1+\gbar^2}} \quad ,
\label{exp-sech}
\eea
and the periodicity condition \eqn{periodic} involves the function
\bea
P(\gbar^2)=\frac{1}{\pi} \int_{-1}^1 \frac{dy}{(1+\gbar^2 \, y^2)\, \sqrt{1-y^2}}=\frac{1}{\sqrt{1+\gbar^2}}\quad .
\label{p-sech}
\eea
\item  For the Minkowski electric field $E(t)=E\, \cos(\omega t)$, we have $f(\omega x_4)=\sinh(\omega x_4)$, so that 
\be
f^\prime(\omega x_4)= \cosh(\omega x_4)= \sqrt{1+\gbar^2\, y^2}\quad .
\ee
Then the exponent \eqn{exponent2} involves the function 
\bea
g(\gbar^2)=\frac{2}{\pi}\int_{-1}^1 \frac{dy \, \sqrt{1-y^2}}{\sqrt{1+\gbar^2 \, y^2}}= \frac{4}{\pi}\frac{\sqrt{1+\gbar^2}}{\gbar^2} \left[ {\bf K}\left(\frac{\gbar^2}{1+\gbar^2}\right)-{\bf E}\left(\frac{\gbar^2}{1+\gbar^2}\right)\right] \quad ,
\label{exp-cos}
\eea
and the periodicity condition \eqn{periodic} involves the function
\bea
P(\gbar^2)=\frac{1}{\pi}\int_{-1}^1 \frac{dy}{\sqrt{1+\gbar^2 \, y^2}\, \sqrt{1-y^2}}=\frac{2}{\pi}\,\frac{1}{\sqrt{1+\gbar^2}}\, {\bf K}\left(\frac{\gbar^2}{1+\gbar^2}\right)\quad .
\label{p-cos}
\eea
\end{itemize}
We identify the critical point of the exponent as follows. From \eqn{exponent2} -- \eqn{periodic} it follows that the derivative with respect to $T$ takes a remarkably simple form:
\bea
\frac{d\Delta(T)}{d T} = m^2\left[1-\frac{\gbar^2(T)}{\gamma^2}\right]\quad .
\label{firstderiv}
\eea
Thus, the critical point $T_c$ of the exponent occurs at $T$ such that $\gbar(T_c)=\gamma$ [as was found before in \eqn{special}], which determines $T_c$ as a particular function of $\gamma$: 
\bea
T_c= \frac{\pi}{e E}\, P(\gamma^2) \quad .
\label{critical}
\eea
Evaluating the exponent at this critical point yields
\bea
\Delta(T_c) = 
\frac{m^2\pi}{e E} g(\gamma^2) \quad ,
\label{criticalexponent}
\eea
which is the exponent derived in \cite{ds}. 
From (\ref{firstderiv}) and (\ref{periodic}), the second derivative of the exponent, evaluated at the critical point, is
\bea
\left[\frac{d^2\Delta(T)}{d T^2}\right]_{T=T_c}= -\frac{m^2 e E}{\pi}\,\frac{1}{\gamma^2 \frac{d  P(\gamma^2)}{d(\gamma^2)} } \quad .
\label{secondderiv}
\eea
The final ingredient for the semiclassical evaluation of the pair production rate is to evaluate the determinant prefactor at $T_c$, for which we find the following simple expression, involving the {\it same} function $P(\gamma^2)$:
\bea
I_1\left( T_c\right) =\left[\int_0^T\frac{d\tau}{(\dot{x}_4^{\rm cl})^2}\right]_{T=T_c}=-\frac{\pi}{2m^2 e E} \, \gamma^2 \,  \frac{d P(\gamma^2)}{d(\gamma^2)} \quad .
\label{i2critical}
\eea

Thus, our final semiclassical approximation for the nonperturbative imaginary part of the Minkowski effective action for a time-dependent electric field background is:
\bea
{\rm Im}\,\Gamma_{\rm Mink}^{\rm semi}&\approx &
 V_3^{\rm Mink} \, \frac{1}{(4\pi)^2}\,\frac{\omega}{4eE}\, \sqrt{\frac{2\pi}{\Delta^{\prime\prime}(T_c)} }  \, \frac{e^{-\Delta(T_c)}}{T_c \gbar(T_c)\, I_1(T_c)}\label{laplace-general}\\
&=& V_3^{\rm Mink} \, \frac{\sqrt{2}(eE)^{5/2}}{32\pi^3m\omega}\, \frac{1}{\frac{d}{d(\gamma^2)} \left(\gamma^2\, g(\gamma^2)\right)\, \sqrt{-\frac{d^2}{d(\gamma^2)^2} \left(\gamma^2\, g(\gamma^2)\right)}}\, \exp\left[-\frac{m^2\pi}{e E}\,g(\gamma^2)\right] \quad ,
\nonumber
\eea
where $g(\gamma^2)$ is the function defined in \eqn{gfunction}, evaluated at $\gbar=\gamma$. Recalling \eqn{pg} that $P(\gamma^2)$ is completely determined by the function $g(\gamma^2)$, we note the important fact that {\it the semiclassical approximation} \eqn{laplace-general} {\it is expressed entirely  in terms of the single function} $g(\gamma^2)$.

For example:
\begin{itemize}
\item 
For $E(t)=E\, {\rm sech}^2(\omega t)$, we find the result in \eqn{laplace-sech}.
\item  For $E(t)=E\, \cos(\omega t)$, we find
\bea
\hskip -1cm {\rm Im}\,\Gamma_{\rm Mink}^{\rm semi} 
&\approx& V_3^{\rm Mink}\,  \frac{ \sqrt{2\pi}\left(e E\right)^{3/2}}{64\pi^2}\, \frac{\left(1+\gamma^2\right)^{3/4}\, \exp\left\{-\frac{4m^2}{eE}\, \frac{\sqrt{1+\gamma^2}}{\gamma^2} \left[ {\bf K}\left(\frac{\gamma^2}{1+\gamma^2}\right)-{\bf E}\left(\frac{\gamma^2}{1+\gamma^2}\right)\right]\right\}}{{\bf K}\left(\frac{\gamma^2}{1+\gamma^2}\right) \sqrt{{\bf K}\left(\frac{\gamma^2}{1+\gamma^2}\right)-{\bf E}\left(\frac{\gamma^2}{1+\gamma^2}\right)}} \,
\,\quad . \nn
\label{semi-cos}
\eea
Once again, in the static limit ($\gamma\to 0$), this reduces to the locally constant field approximation result
\bea
 {\rm Im}\,\Gamma_{\rm Mink}^{\rm LCF} 
&\approx& V_3^{\rm Mink}\, 
\frac{e^2}{16\pi^3}\int_{-\frac{\pi}{2\omega}}^{\frac{\pi}{2\omega}} dt\, E^2\cos^2(\omega t)
\,\exp\left[-\frac{m^2 \pi}{e E | \cos(\omega t)|}\right]\nn
&\sim &V_3^{\rm Mink}\,  \frac{\sqrt{2}\left(e E\right)^{5/2}}{16\pi^3 m\omega}
\,\exp\left[-\frac{m^2 \pi}{e E}\right]\quad .
\label{lcf-cos}
\eea
The result \eqn{semi-cos} also agrees with Popov's WKB analysis, after some (presumably typographical) errors are corrected in \cite{popov}.
\end{itemize}

We stress the simplicity and versatility of the result \eqn{laplace-general}.
It means that for background fields where the gauge field (and hence the electric field) is a function of just one space-time coordinate, one does not even have to find the explicit form of the semiclassical worldline instanton path which dominates the functional integral. Instead, one simply needs to compute the function $g(\gamma^2)$ [defined in (\ref{gfunction})], and its first few derivatives. Even if this cannot be done in closed-form, it could be done numerically. As a final example we can take the time-dependent Minkowski electric field
\bea
E(t)=\frac{E}{\left(1+\left(\omega t\right)^2\right)^{3/2}} \quad ,
\label{eg}
\eea
for which the Euclidean gauge function is
\bea
f(\omega x_4)= \frac{\omega x_4}{\sqrt{1-\left(\omega x_4\right)^2}} \quad .
\eea
Then the function $g(\gamma^2)$ in \eqn{gfunction} is
\be
g(\gamma^2)= \frac{2}{\pi} \int_{-1}^1\, dy \,\frac{\sqrt{1-y^2}}{\left(1+\gamma^2 y^2\right)^{3/2}}
=\frac{4}{\pi\, \gamma^2} \left[ {\bf E}(-\gamma^2) - {\bf K}(-\gamma^2) \right]\quad. 
\ee
Then the semiclassical imaginary part of the Minkowski effective action is 
\bea
{\rm Im}\,\Gamma_{\rm Mink}^{\rm semi} 
&\approx& V_3^{\rm Mink}\,  \frac{\sqrt{2\pi}\left(e E\right)^{3/2}}{64\pi^2}\, \frac{\left(1+\gamma^2\right)^{3/4}\, \exp\left\{-\frac{4m^2}{e E \gamma^2 } \left[ {\bf E}\left(-\gamma^2\right)-{\bf K}\left(-\gamma^2\right)\right]\right\}}{{\bf E}\left(-\gamma^2\right) \sqrt{(1+\gamma^2) {\bf K}\left(-\gamma^2\right)-(1-\gamma^2){\bf E}\left(-\gamma^2\right)}}
\nn
\label{semi-eg}
\eea
Note that while we could have computed the explicit worldline instanton path for this background, it was in fact not necessary in order to compute the semiclassical imaginary part of the Minkowski effective action. In the static limit ($\gamma\to 0$), this reduces to the locally constant field approximation result
\bea
 {\rm Im}\,\Gamma_{\rm Mink}^{\rm LCF} 
&\approx& V_3^{\rm Mink}\, 
\frac{e^2}{16\pi^3}\int_{-\infty}^{\infty} dt\, \frac{E^2}{\left[1+(\omega t)^2\right]^3}
\,\exp\left\{-\frac{m^2 \pi}{e E}\left[1+(\omega t)^2\right]^{3/2}\right\}\nn
&\sim &V_3^{\rm Mink}\,  \frac{\sqrt{2}\left(e E\right)^{5/2}}{16\pi^3\sqrt{3} \, m\omega}
\,\exp\left[-\frac{m^2 \pi}{e E}\right]\quad .
\label{lcf-eg}
\eea
The result \eqn{semi-eg} also agrees with Popov's WKB analysis, after some (presumably typographical) errors are corrected in \cite{popov}.

\section{Spatially Inhomogeneous Electric Fields}
\label{space}

As discussed already in \cite{ds}, spatially inhomogeneous electric fields that are functions of a single spatial coordinate, say $x_3$, can be treated analogously. Consider the class of spatially inhomogeneous electric fields with Euclidean gauge field
\bea
A_4(x_3)=-i \frac{E}{k}\, f(k x_3) \quad.
\label{xgauge}
\eea
For example, the single-bump Minkowski electric field $E(x_3)=E\, {\rm sech}^2(k x_3)$ has $f(k x_3)=\tanh(k x_3)$, while the periodic Minkowski electric field $E(x_3)=E\, \cos(k x_3)$ has $f(k x_3)=\sin(k x_3)$. Define the {\it spatial adiabaticity parameter}
\bea
\tilde{\gamma}=\frac{m k}{e E}\quad.
\label{gt}
\eea
The entire analysis of worldline instantons can be repeated as in the time-dependent case of  Section \ref{time}, although the $T$ integral requires an analytic continuation to the complex plane in order to be evaluated by steepest descents. In fact, the final results for the imaginary part of the Minkowski effective action can be obtained from those of the corresponding time-dependent system by the analytic continuation:
\bea
\gamma\to i\, \tilde{\gamma}\quad.
\label{continuation}
\eea
Thus, our  semiclassical approximation for the nonperturbative imaginary part of the Minkowski effective action for a space-dependent electric field background with gauge function \eqn{xgauge} is:
\bea
{\rm Im}\,\Gamma_{\rm Mink}^{\rm semi}\approx 
 \left(V_2 {\mathcal T}\right)^{\rm Mink}  \frac{\sqrt{2}(eE)^{5/2}}{32 \pi^3m \, k} \frac{\exp\left[-\frac{m^2\pi}{e E}\,\tilde{g}(\tilde{\gamma}^2)\right] }{\frac{d}{d(\tilde{\gamma}^2)} \left(\tilde{\gamma}^2\, \tilde{g}(\tilde{\gamma}^2)\right)\, \sqrt{\frac{d^2}{d(\tilde{\gamma}^2)^2} \left(\tilde{\gamma}^2\, \tilde{g}(\tilde{\gamma}^2)\right)}}\, \quad ,
\label{space-general}
\eea
where $\tilde{g}(\tilde{\gamma}^2)$ is the function defined by
\bea
\tilde{g}(\tilde{\gamma}^2)= \frac{2}{\pi} \int_{-1}^1\, dy \,\frac{\sqrt{1-y^2}}{|f^\prime|}\quad,
\label{space-g}
\eea
where we have written $y=\frac{f(k \, x_4^{\rm cl})}{\tilde{\gamma}}$, and $f^\prime$ means the derivative is re-expressed as a function of $y$. For example:
\begin{itemize}
\item For the Minkowski electric field $E(x_3)=E\, {\rm sech}^2(k x_3)$, we have $f(k x_3)=\tanh(k x_3)$, and
$f^\prime(k x_3)= {\rm sech}^2(k x_3)= 1-\tilde{\gamma}^2\, y^2$. Thus, 
\bea
\tilde{g}(\tilde{\gamma}^2)=\frac{2}{\pi}\int_{-1}^1 \frac{dy \, \sqrt{1-y^2}}{(1-\tilde{\gamma}^2 \, y^2)}=\frac{2}{1+\sqrt{1-\tilde{\gamma}^2}} \quad .
\label{gt-sech}
\eea
The imaginary part of the Minkowski effective action is 
\bea
{\rm Im}\,\Gamma_{\rm Mink}^{\rm semi}
&=&  \left(V_2 {\mathcal T}\right)^{\rm Mink}  \, \frac{\left(e E\right)^{5/2}}{16\pi^3m k} \left(1-\tilde{\gamma}^2\right)^{5/4} \, \exp\left[-\frac{m^2\pi}{e E} \, \left(\frac{2}{1+\sqrt{1-\tilde{\gamma}^2}} \right) \right]\,\, .
\label{x-sech}
\eea
The corresponding locally constant field approximation is
\bea
{\rm Im}\,\Gamma_{\rm Mink}^{\rm LCF}
&=&  \left(V_2 {\mathcal T}\right)^{\rm Mink} \, \frac{e^2 E^2}{16\pi^3}\, \int_{-\infty}^\infty dx_3\, {\rm sech}^4(k x_3)\, \exp\left[-\frac{m^2\pi}{e E}\,\cosh^2(k x_3)\right]\nn
&\sim& \left(V_2 {\mathcal T}\right)^{\rm Mink} \, \frac{\left(e E\right)^{5/2}}{16\pi^3m k} \, \exp\left[-\frac{m^2\pi}{e E}  \right]\quad .
\label{x-sech-lcf}
\eea
The ratio of the semiclassical answer \eqn{x-sech} to the LCF approximation \eqn{x-sech-lcf} is plotted in Figure \ref{fig1}, along with the exact result of \cite{nikishov} and the numerical results of \cite{giesklingmuller}. The agreement between the semiclassical expression and the numerical results is excellent. This is especially true given that the numerical data is from a system with $\frac{e E}{m^2}=1$, which is far from the weak field limit in which the semiclassical expression was derived. We also comment that \eqn{space-general} and \eqn{x-sech} agree with the quantum mechanical instanton result of Kim and Page \cite{kimpage,page}, after a gaussian integration over energy and momentum.
\begin{figure}[tb]
\begin{center}
\includegraphics[width=0.8\textwidth]{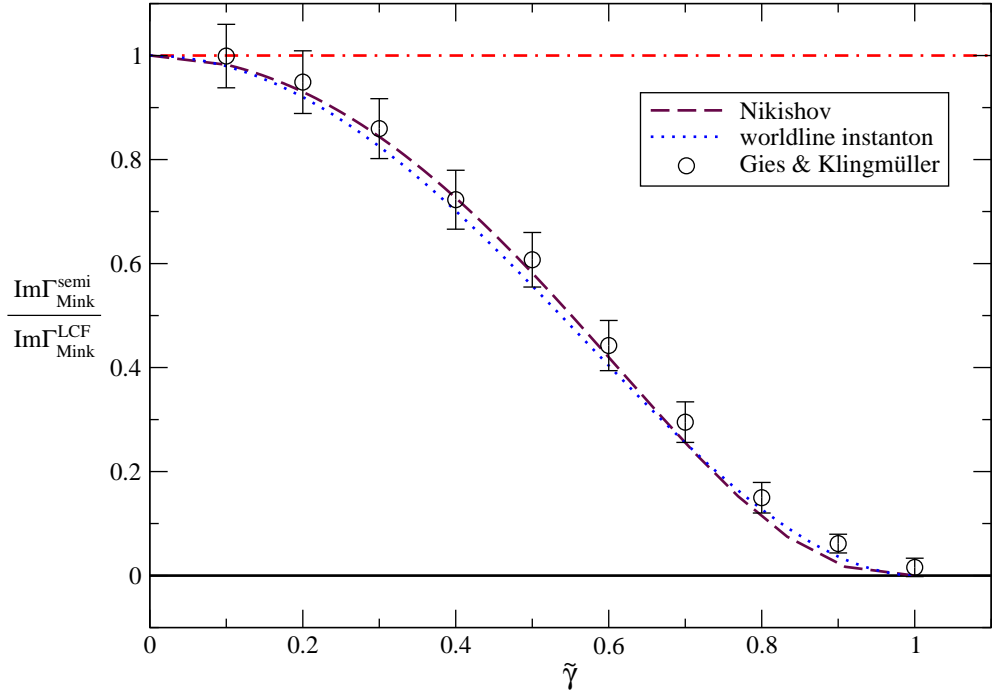}
\end{center}
\caption{The dotted line plots the ratio of our semiclassical worldline instanton expression \eqn{x-sech} to the weak field limit of the corresponding locally constant field approximation \eqn{x-sech-lcf}. The dashed line is the same ratio using a numerical integration of the exact expression, derived from Nikishov's exact result in \cite{nikishov} (see also \cite{kimpage}). The circles represent the numerical worldline results of Gies and Klingm\"uller \cite{giesklingmuller}, which were evaluated for $\frac{e E}{m^2}=1$. Note that the agreement is excellent, even though it is far from the weak field limit. }
\label{fig1}
\end{figure}

\item  For the Minkowski electric field $E(x_3)=E\, \cos(k x_3)$, we have $f(k\, x_3)=\sin(k\, x_3)$, and 
$f^\prime(k x_3)= \cos(k\,x_3)=\sqrt{1-\tilde{\gamma}^2\, y^2}$.
Thus
\bea
\tilde{g}(\tilde{\gamma}^2)=\frac{2}{\pi}\int_{-1}^1 \frac{dy \, \sqrt{1-y^2}}{\sqrt{1-\tilde{\gamma}^2 \, y^2}}=  \frac{4\sqrt{1-\tilde{\gamma}^2}}{\pi\tilde{\gamma}^2} \left[ {\bf E}\left(\frac{-\tilde{\gamma}^2}{1-\tilde{\gamma}^2}\right)-{\bf K}\left(\frac{-\tilde{\gamma}^2}{1-\tilde{\gamma}^2}\right)\right] \quad .
\label{gt-cos}
\eea
The imaginary part of the Minkowski effective action is 
\bea
\hskip -1cm {\rm Im}\,\Gamma_{\rm Mink}^{\rm semi} 
&\approx&  \left(V_2 {\mathcal T}\right)^{\rm Mink}\,  \frac{ \sqrt{2\pi}\left(e E\right)^{3/2}}{64\pi^2}\, \frac{\left(1-\tilde{\gamma}^2\right)^{3/4}\, \exp\left\{-\frac{4m^2}{eE}\, \frac{\sqrt{1-\tilde{\gamma}^2}}{\tilde{\gamma}^2} \left[ {\bf E}\left(\frac{-\tilde{\gamma}^2}{1-\tilde{\gamma}^2}\right)-{\bf K}\left(\frac{-\tilde{\gamma}^2}{1-\tilde{\gamma}^2}\right)\right]\right\}}{{\bf K}\left(\frac{-\tilde{\gamma}^2}{1-\tilde{\gamma}^2}\right) \sqrt{{\bf E}\left(\frac{-\tilde{\gamma}^2}{1-\tilde{\gamma}^2}\right)-{\bf K}\left(\frac{-\tilde{\gamma}^2}{1-\tilde{\gamma}^2}\right)}} \,
\,\quad . \nn
\label{xsemi-cos}
\eea
\item
For the Minkowski electric field $E(x_3)=\frac{E}{\left[1+\left(k x_3\right)^2\right] ^{3/2}}$, we have 
$f(k x_3)= \frac{k x_3}{\sqrt{1+\left(k x_3\right)^2}}$, and $f^\prime(k x_3)=\left(1-\tilde{\gamma}^2 y^2\right)^{3/2}$. 
Thus,
\be
\tilde{g}(\tilde{\gamma}^2)=\frac{2}{\pi} \int_{-1}^1\, dy \,\frac{\sqrt{1-y^2}}{\left(1-\tilde{\gamma}^2 y^2\right)^{3/2}}= \frac{4}{\pi \tilde{\gamma}^2} \,\left[ {\bf K}\left(\tilde{\gamma}^2\right) - {\bf E}\left(\tilde{\gamma}^2\right)\right]\quad.
\ee
Then the semiclassical imaginary part of the Minkowski effective action is
\bea
{\rm Im}\,\Gamma_{\rm Mink}^{\rm semi} 
&\approx& \left(V_2 {\mathcal T}\right)^{\rm Mink}\,  \frac{\sqrt{2\pi}\left(e E\right)^{3/2}}{64\pi^2}\, \frac{\left(1-\tilde{\gamma}^2\right)^{3/4}\, \exp\left\{-\frac{4m^2}{e E \tilde{\gamma}^2 } \left[ {\bf K}\left(\tilde{\gamma}^2\right)-{\bf E}\left(\tilde{\gamma}^2\right)\right]\right\}}{{\bf E}\left(\tilde{\gamma}^2\right) \sqrt{(1+\tilde{\gamma}^2) {\bf E}\left(\tilde{\gamma}^2\right)-(1-\tilde{\gamma}^2){\bf K}\left(\tilde{\gamma}^2\right)}}
\nn
\label{xsemi-eg}
\eea
Note that while we could have computed the explicit worldline instanton path for this background, it was in fact not necessary in order to compute the semiclassical imaginary part of the Minkowski effective action. 

\end{itemize}

\section{Conclusions}
\label{conclusions}

To conclude, the {\it worldline instanton} approach has now been
extended to include also the quantum fluctuation prefactor for the
nonperturbative imaginary part of the effective action. For general
background fields the computation is numerical. Given the numerically
determined instanton loop, the fluctuation determinant can be computed
directly using \eqn{semi}, which is a result from the semiclassical
analysis of quantum mechanical path integrals \cite{levit,kleinertbook}. For
inhomogeneous time-dependent electric fields, the analysis can be done
in much more explicit form, culminating in the semiclassical
expression \eqn{laplace-general}, which is expressed entirely in terms
of the function $g(\gamma^2)$ defined in \eqn{gfunction}. Similarly,
for inhomogeneous space-dependent electric fields, the corresponding
expression is \eqn{space-general}, with $\tilde{g}(\tilde{\gamma}^2)$ defined
in \eqn{space-g}. The agreement with Popov's WKB analysis \cite{popov}
is perfect, and the semiclassical results match the numerical results
of \cite{giesklingmuller} very well. Within the semiclassical approximation it appears
that the existence of a worldline instanton ({\it i.e.}, the existence of a periodic
solution to the classical Euclidean equations of motion) is a signal
for the existence of an imaginary part to the Minkowski effective action, and hence
for particle production. 
Since we are working in the semiclassical approximation we cannot necessarily 
  conclude that, conversely, the absence of a worldline instanton solution would
  imply the absence of pair creation. Nevertheless, it is interesting to
  observe that, in the case of the spatially inhomogeneous electric field $E(x_3) =
  E\, {\rm sech}^2(k x_3)$ treated in section IV, there is no periodic worldline instanton
when  $\tilde{\gamma} >1$, and this is precisely the regime in which
the imaginary part of the effective action vanishes, {\sl even away from the weak field limit} \cite{nikishov,giesklingmuller}. Similarly, it is easy to see from the classical equations of motion (\ref{euler}) that there is no
worldline instanton for a plane-wave  background field, consistent with the absence of 
pair production in this case \cite{schwinger}.
A deeper physical and geometrical understanding of this correspondence 
would be interesting.

A number of important issues remain. First, while the agreement
between our final answer and the worldline numerical approach
are very good, the details of the calculation are very different. In
this computation the nonperturbative result comes from small quantum
fluctuations around a single closed loop amongst the ensemble of all
closed loops. On the other hand, the worldline numerics does not
appear to be dominated by single loops. For the electric field
configurations considered here, the dominance of worldline instantons
could directly be tested, for instance, by reweighting the numerical
worldline ensemble with the instanton configurations. In turn, a
cooling procedure on top of the worldline numerical algorithm can be
used to numerically determine the instanton configurations needed for
the present approach if applied to more complicated background
fields.  A better understanding of this correspondence should lead to
more efficient numerical worldline loop computations, and should also
clarify the physical nature of the semiclassical approximation for
more general types of background field. This relates to the fact that
the standard WKB approaches of Br\'ezin and Itzykson \cite{brezin},
Popov {\it et al} \cite{popov}, and Kim and Page \cite{kimpage}, are difficult to
generalize to more complicated fields as multidimensional WKB is
considerably more difficult than one-dimensional WKB. On the other
hand, the worldline fluctuation problem is inherently one-dimensional,
once the worldline instanton loop has been found. 

Second, the results
here are for scalar QED. The generalization to spinor QED has been
explained in \cite{ds} for the cases where the electric background is
a function of just one spacetime coordinate, in which case the spin
factor reduces to a factor of $-2\,(-1)^n$ inside the sum over
multi-instantons of instanton number $n$. So for the leading
single-instanton piece, the modification is simply the spin degeneracy
factor of $2$. However, for more complicated background fields, it is
not clear how to evaluate the spin factor efficiently. This will be
addressed in future work. Third, this paper has considered QED. The
worldline expression \eqn{effective} can be generalized also to
nonabelian gauge theories \cite{schmidt,csreview}. In this case the
semiclassical worldline loops would then be related to Wong's
equations \cite{wong}. This, in turn, may be useful for applications 
to the color glass condensate \cite{raju}. Finally, one could use the semiclassical
approximation to address higher loop effects, using the higher loop
worldline formalism for effective actions \cite{schmidt}. Thus the
worldline instanton approach has the potential to address higher
loops, while it is not at all clear how to address higher loops in the
WKB language. The main result of Affleck {\it et al}'s work \cite{affleck}
is that for a constant $E$ field, the instanton approach provides a
way to resum the leading effect of all higher loops in the situation
where the constant field $E$ is weak, but the coupling $e$ is
arbitrary. This is because the instanton solution remains a stationary
point even after taking the additional interaction term into account
which in the worldline formalism represents virtual photon exchanges
in the loop. It would be very interesting to try to extend this type
of analysis to the general worldline instanton loops for inhomogeneous
background fields.  This should make contact with the work of
Halpern {\it et al}, who considered a new type of strong-coupling
expansion in the worldline approach \cite{halpern}.

\vskip 1cm

{\bf Acknowledgements:} We are very grateful to Don Page  for helpful comments and correspondence. GD and QW thank the US DOE for support through the grant DE-FG02-92ER40716. GD, CS and QW acknowledge the support of the NSF US-Mexico Collaborative Research Grant 0122615. HG acknowledges support by the DFG under contract Gi 328/1-3
(Emmy-Noether program) and Gi 328/3-2.

\end{document}